%% file: main.tex
\documentclass[submission,copyright,creativecommons]{eptcs}
\providecommand{\event}{FMAS 2025} 

\usepackage{iftex}
\usepackage{amsmath,amssymb}
\usepackage{siunitx}
\usepackage[section]{placeins}
\usepackage{acronym}
\usepackage{graphicx}
\usepackage{enumitem}
\usepackage[cal=cm]{mathalpha}
\usepackage{color,soul}
\usepackage{algorithm}
\usepackage{algpseudocodex}
\usepackage{booktabs}
\usepackage[font=small,labelfont=bf]{caption}
\usepackage{subcaption}
\ifpdf
  \usepackage{underscore}         
  \usepackage[T1]{fontenc}        
\else
  \usepackage{breakurl}           
\fi

\newcommand{\reffig}[1]{Fig.~\ref{fig:#1}}

\newcommand{\refsec}[1]{Sec.~\ref{sec:#1}}

\newcommand{\refalg}[1]{Alg.~\ref{alg:#1}}

\title{Abstract Scene Graphs: Formalizing and Monitoring Spatial Properties of Automated Driving Functions}
\author{Ishan Saxena \qquad\qquad Bernd Westphal
\institute{
German Aerospace Center (DLR)\\
Institute of Systems Engineering for Future Mobility\\
Oldenburg, Germany}
\email{\quad ishan.saxena@dlr.de \quad\quad bernd.westphal@dlr.de}
\and
Martin Fränzle
\institute{Carl von Ossietzky University of Oldenburg\\
Oldenburg, Germany}
\email{martin.fraenzle@uni-oldenburg.de}
}
\def\titlerunning{Abstract Scene Graphs: Formalizing and Monitoring Spatial Properties of ADFs}
\def\authorrunning{I. Saxena, B. Westphal \& M. Fränzle}
\begin{document}
\maketitle
\vspace{-1em}
\begin{abstract}
Automated Driving Functions~(ADFs) need to comply with spatial properties of varied complexity while driving on public roads. 
Since such situations are safety-critical in nature, it is necessary to continuously check ADFs for compliance with their spatial properties.
Due to their complexity, such spatial properties need to be formalized to enable their automated checking. 
Scene Graphs~(SGs) allow for an explicit structured representation of objects present in a traffic scene and their spatial relationships to each other. 
In this paper, we build upon the SG construct and propose the Abstract Scene Graph~(ASG) formalism to formalize spatial properties of ADFs.
We show using real-world examples how spatial properties can be formalized using ASGs.
Finally, we present a framework that uses ASGs to perform Runtime Monitoring of ADFs. To this end, we also show algorithmically how a spatial property formalized as an ASG can be satisfied by ADF system behaviour.  

\end{abstract}
\thanks{\footnotesize{\noindent%
The research leading to these results is funded by the German Federal Ministry of Education and Research under grant agreement No~16MEE044 (EdgeAI-Trust) and
by the Chips Joint Undertaking under grant agreement No~101139892 (EdgeAI-Trust).}}
\input{sections/intro.tex}
\input{sections/asgs.tex}

\input{sections/nhtsa\_scenarios.tex}

\input{sections/rm\_asg.tex}


\input{sections/conclusion.tex}

\bibliographystyle{eptcs}
\bibliography{main}
\newpage
\appendix
\input{sections/appendix.tex}
\end{document}

%% file: sections/intro.tex
\section{Introduction}
\label{sec:intro}



Ensuring safe operation of Automated Driving Functions~(ADFs) in the automotive domain is a challenging task. 
The complex spatial properties occurring in the automotive domain due to the presence of varied road infrastructure, multitude of traffic participants, and complex traffic rules and social norms presents challenges during ADF development.
Despite scoring high on benchmarks, testing of publicly available ADFs shows that they still violate their safety properties~\cite{Toledo2024}. 
Hence, there is a need to safeguard such ADF driven systems during their operation by checking continuously whether they are satisfying their spatial properties. 
In order to perform such a check, Runtime Monitoring~(RM) of ADFs requires the formalization of such complex spatial properties. 
Scene Graphs~(SGs) are a structured representation that are used to describe objects present in a scene and their spatial relationships with each other. 
The use of SGs, especially for automated driving, has increased quite rapidly of late. 
They have been applied to a variety of use-cases  including scene understanding~\cite{Li2023, Monninger2023, Wang2023}, risk assessment~\cite{Li2021, Malawade2022, Yu2020}, scenario coverage~\cite{Woodlief2024}, motion \& trajectory prediction~\cite{Zipfl2022, Zipfl2023}, test-case reduction~\cite{Zipfl2023a} and monitoring of safety properties~\cite{Toledo2024}.

In this work, we provide a research preview on formalization of spatial properties of ADFs present in the automotive domain based on Scene Graphs.
Concretely, we propose the Abstract Scene Graph~(ASG) formalism.
This formalism takes advantage of the visual structured scene representation of SGs and extends it to enable expression of desired spatial properties of ADFs.
This work builds upon~\cite{Saxena2025} by providing extended formal description and syntax details about ASGs, new examples of real-world, complex spatial properties formalized using ASGs and an algorithmic implementation for checking satisfaction of spatial properties specified as ASGs.

Formalization of spatial properties is a well researched topic in literature.
There exist multiple formalisms that focus specifically on spatial properties occurring in the automotive domain, and a number of them have also been used for performing RM.
MLSL~\cite{Hilscher2011,Schwammberger2021} is a popular formalization language for expressing spatial properties on multi-lane motorways, urban scenarios and country roads and analyse controllers of automated driving systems for safety. 
An offline monitoring concept for MLSL was presented in~\cite{Ody2017}.
The Traffic Sequence Chart~(TSC) formalism~\cite{Damm2018,Damm2018a} is a visual spatio-temporal specification language for specifying system properties in multiple transportation domains. 
TSCs provide the Spatial View construct for specifying spatial properties visually. RM for TSCs has been presented in~\cite{Grundt2022,Stemmer2025}.
The $\mathcal{MS}$~\cite{Kutz2003} logic has also been used to specify spatial properties of autonomous vehicles and perform runtime monitoring by \cite{Pedro2024}. 
The spatial logic $\mathcal{S}4_{{u}}$ combined with Signal Temporal Logic was used for formalizing desired behaviour of cyber-physical systems and performing RM in~\cite{Li2020,Li2021a}
However, the formalisms listed above specify spatial relations between various traffic participants and infrastructure implicitly, e.g., using geometric relations to specify the presence of a vehicle on a particular lane. 
They also lack a suitable representation for handling the varied geometry of road traffic infrastructure, such as curves, and the resultant spatial relations. 
Finally, in order to perform RM, these methodologies require a continuous pre-processing and mapping step for the captured sensor signals to enable evaluation of predicate-based atomic formula, increasing the computational cost.

Closest to our approach, in \cite{Toledo2024,Woodlief2025a} the authors introduce frameworks for performing RM of spatial and temporal safety properties of autonomous systems based on the Scene Graph Language and \textsc{SceneFlow} domain-specific languages. 
They show that traffic scene data captured as Scene Graphs is useful for performing RM. 
However, the framework requires additional domain knowledge from ADF developers to formalize text-based safety properties using a textual graph query language.
This could lead to the introduction of undesired errors and hinders discussion about the formalization with other stakeholders. 
Further, they have not shown how their framework can be used for specifying and monitoring other types of properties, e.g. behavioural, of autonomous systems.

\textbf{Outline.} The rest of the paper is organized as follows.
\refsec{asgs} introduces the Abstract Scene Graph~(ASG) formalism. 
In \refsec{nhtsa}, we show how two example spatial properties from the NHTSA pre-crash scenarios catalogue~\cite{Najm2007} can be formalized using ASGs.
A concept for performing Runtime Monitoring using ASGs is presented in \refsec{rmasg}.  
We conclude this paper and provide details about future work in \refsec{conclusion}

%% file: sections/asgs.tex
\section{Abstract Scene Graphs}
\label{sec:asgs}
Scene Graphs~(SGs) are used to represent objects present in a specified area of interest, objects' attributes and spatial relations between objects as a collection of nodes and edges in a graph or tree structure.
Abstract Scene Graphs~(ASGs) build upon SGs and enable formal description of desired functionality of objects and their spatial behaviour.
Concretely, ASGs restrict the set of scenes represented by an SG to those consisting of the desired objects, their corresponding spatial relations and attribute value range.

\textbf{Preliminaries:} We now introduce terminology and definitions, which we will use later for defining ASGs.
First, we assume a \textit{traffic scene} in the automotive domain to consist of traffic participants and traffic environment.
A \textit{traffic participant} is an entity that interacts inside a traffic environment and with other traffic participants according to a set of physical and behavioural rules. 
A traffic participant may also contain attributes such as position, velocity, etc. Examples include vehicles, pedestrians, etc.
A \textit{traffic environment} is an entity that represents the setting in which traffic participants interact, e.g. roads, lanes, etc.
We further define an \textit{Object Model}, to model traffic participants and traffic environment. 
It is represented as $OM = (\mathcal{C}, \mathcal{T}, \mathcal{P}, \mathcal{F}, \mathcal{E})$, where $\mathcal{C}$ is the set of object classes, $\mathcal{T}$ is a set of basic types and $attr(\mathcal{C})$ is the finite set of typed attributes for each object class. $\mathcal{P}$ is the set of typed predicate symbols, $\mathcal{F}$ is the set of typed function symbols and $\mathcal{E}$ is the set of relationships allowed between the object classes present in $\mathcal{C}$.
A \textit{concrete traffic scene} consists of a snapshot of traffic participants and traffic environment present in traffic scene, along with their attributes and relations to each other~\cite{Ulbrich2015}.
Formally, they are defined, over $OM$, as $\sigma: \text{ID}\nrightarrow (attr(\mathcal{C})\nrightarrow \mathcal{D})$, where ID is the finite set of object identities corresponding to objects present in the scene and $\mathcal{D}$ is the domain set of attribute types.
Finally, a directed heterogeneous graph is a directed graph structure consisting of multiple types of nodes and edges. It also allows multiple edges to exist between the same pair of nodes. 

\begin{figure}[t]
	\centering
	\includegraphics[width=0.4\linewidth]{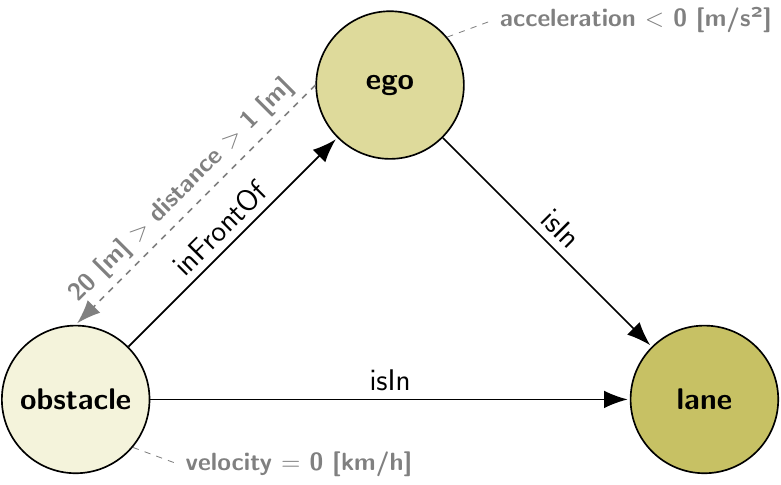}
	\caption{An example Abstract Scene Graph~\cite{Saxena2025}}
    \label{fig:asgeg}
\end{figure}

ASGs aim to formally describe desired state and spatial properties of traffic scenes. 
They should be interpreted as a set of well typed predicate logic formulae over $OM$. 
An ASG is represented as $ASG: (G_{\text{A}}, D)$, where $G_{\text{A}}$ is a directed heterogeneous graph and $D$ is a set of predicates pertaining to graph elements. 
Semantically, $G_{\text{A}}$ represents a traffic scene with traffic participants and traffic environment encoded as nodes and the spatial relationships between them encoded as edges.
The set $D$ represents the desired attribute values of traffic participants/environment elements which are encoded as predicates.
The node and edge types present in $G_{\text{A}}$ are a subset of those present in $OM$.
An example ASG expressing the property "ego-vehicle starts breaking once it is at most \qty{20}{\metre} in front of the static obstacle" is shown in~\reffig{asgeg}.
Object nodes are labelled with their identities while relation types are visualized by labels on top of edges.
The elements of predicate set $D$ are visualized using grey colour in \reffig{asgeg}. 
The predicates constraining attributes of a single element are directly attached to the corresponding graph element. E.g., the text attached to the obstacle node constrains its velocity and the corresponding predicate logic formula should be interpreted as $obstacle.velocity=0$.
Predicates can additionally also constrain properties that exist between a pair of object nodes. E.g., in~\reffig{asgeg} the dashed edge constrains the distance between the ego and obstacle objects to a certain range.

Given a suitable $OM$, ASGs can be used to express a variety of spatial properties from different domains. 
But, in this paper, we restrict the discussion to spatial properties arising from the Scenario-based Testing~\cite{Neurohr2020} of Automated Driving Functions in the automotive domain.
Specifying spatial properties using ASGs provides the benefit that it allows for explicit modelling of the traffic scene and reduces semantic ambiguity in spatial reasoning.
In addition, ASGs enable multi-stakeholder discussion of spatial rules and regulations, since their usage doesn't require specialized domain knowledge.   






%% file: sections/nhtsa_scenarios.tex
\section{Formalization of Spatial Properties using Abstract Scene Graphs}
\label{sec:nhtsa}
In this section, we provide details about how spatial properties may be formalized as Abstract Scene Graphs~(ASGs).
Specifically, we provide examples of two properties present in the NHTSA pre-crash scenario catalogue~\cite{Najm2007}.
These properties are provided as scenario descriptions in natural language, which are known as functional scenario description in Scenario-based Testing literature~\cite{Menzel2018,Neurohr2025}. 
However, such descriptions are unsuitable for performing automatic checks to identify whether an Automated Driving Function~(ADF) is satisfying its properties. 
They should first be formalized into an objective, machine-readable format. 
We now present this formalization process for ASGs. 
For both scenarios, we assume that the ADF is driving on a two-lane road inside city boundaries with a speed limit of \qty{30}{\kilo\metre\per\hour}.  

\subsection{Example I - Parking Exit}
\textbf{Property P1: }\textit{The ego-vehicle must exit a parallel parking bay safely into a flow of traffic.}

\begin{figure*}[t!]
    \centering
    \begin{subfigure}[t]{0.47\linewidth}
        \centering
        \includegraphics[width=0.8\columnwidth]{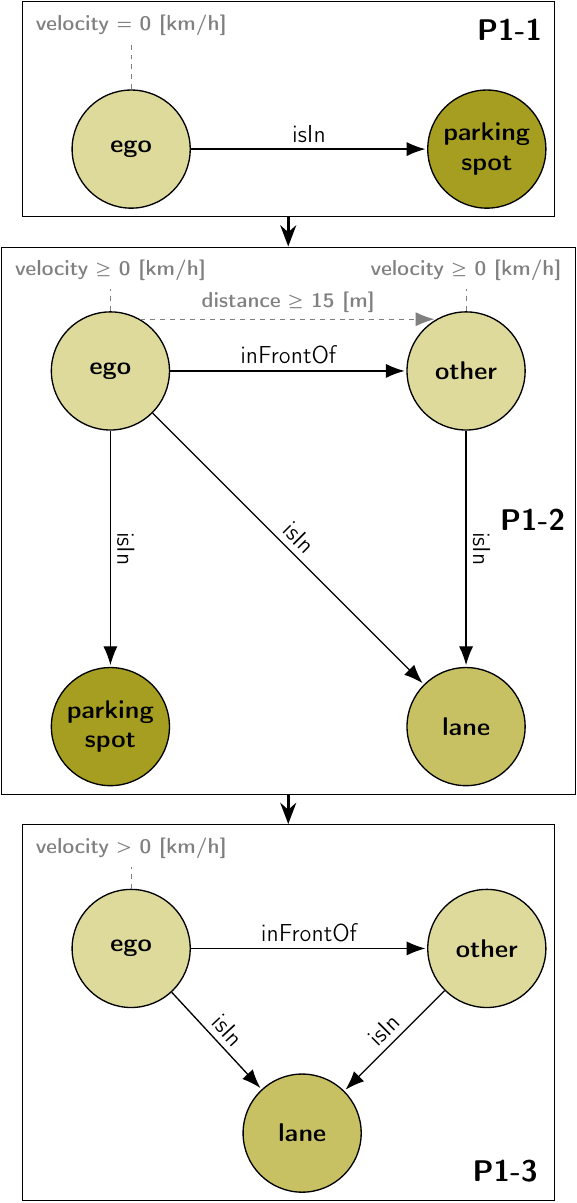}
        \caption{\textbf{P1}: Parking Exit}
		\label{fig:p2asg}
    \end{subfigure}%
    ~
    \begin{subfigure}[t]{0.52\linewidth}
        \centering
        \includegraphics[width=\columnwidth]{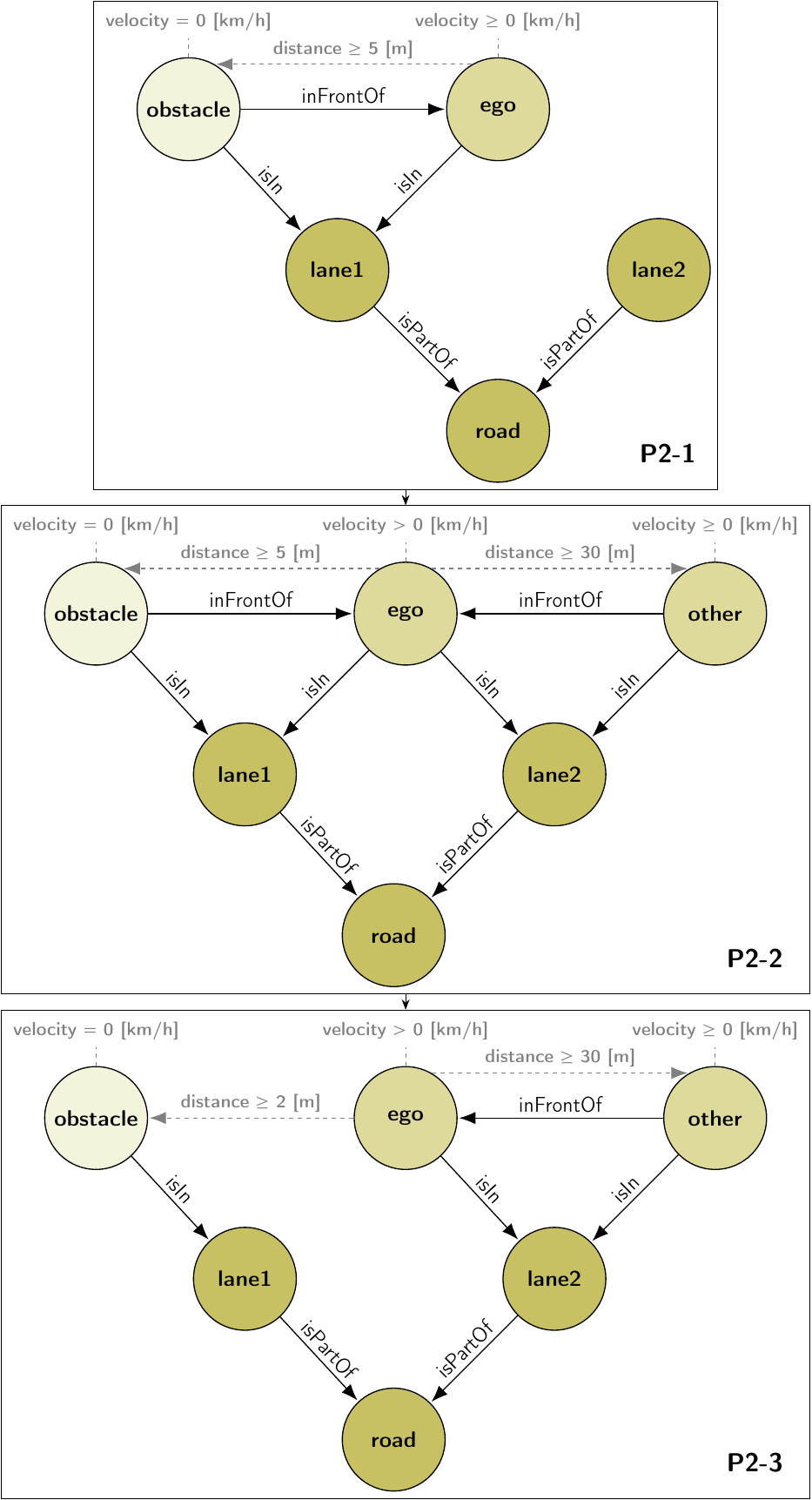}
        \caption{\textbf{P2}: Obstacle in Lane}
		\label{fig:obsasg}
    \end{subfigure}
    \caption{Abstract Scene Graphs for example spatial properties}
\end{figure*}

In \textbf{P1}, the behaviour of two traffic participants ego vehicle and oncoming traffic as well as two traffic environment entities a driving lane and a parking spot is described. We model these objects using $Vehicle$, $Lane$ and $ParkingSpot$ object types. The functional description of \textbf{P1} can be understood to be composed of the following phases:  

\begin{itemize}
    \item[P1-1] ego is standing still in the parking spot 
    \item[P1-2] ego is turning into the driving lane next to the parking spot while maintaining a safe distance (at least \qty{15}{\metre}) to the next vehicle coming from behind 
    \item[P1-3] ego has joined the lane completely and is driving ahead 
\end{itemize}
The ASGs corresponding to the phase descriptions for \textbf{P1} are shown in~\reffig{p2asg}.
\subsection{Example II - Obstacle in Lane}
\textbf{Property P2: }\textit{The ego-vehicle encounters a static obstacle blocking the lane and must perform a lane change into traffic moving in the opposite direction to avoid it. It should then return to its original lane.}

\textbf{P2} describes the behaviour of six distinct objects: an ego vehicle, two lanes, a road that contains both lanes, a static obstacle and oncoming traffic. 
These objects are modelled by four object types $Vehicle$, $Static$, $Lane$ and $Road$. 
Based on the description provided in \textbf{P2}, we identify five phases that the ego vehicle would need to follow: 
\begin{itemize}
    \item[P2-1] ego is approaching an obstacle present in its lane and maintains a safe distance of at least \qty{5}{\metre} from the obstacle 
    \item[P2-2] ego is changing its lane while maintaining safe distance to the oncoming vehicle (at least \qty{30}{\metre}) and obstacle (at least \qty{5}{\metre})  
    \item[P2-3] ego is completely on the oncoming lane and moving ahead to pass-by the obstacle and maintains safe distance to both the oncoming vehicle (at least \qty{30}{\metre}) and obstacle (at least \qty{2}{\metre}) 
    \item[P2-4] ego has fully passed by the obstacle and is at least \qty{5}{\metre} ahead and starts moving back to its original lane while still maintaining a safe distance to the oncoming vehicle (at least \qty{20}{\metre}) 
    \item[P2-5] ego is completely on its original lane and continues driving ahead 
\end{itemize}
The ASGs corresponding to the first three phase descriptions are shown in~\reffig{obsasg}.

%% file: sections/rm_asg.tex
\section{Runtime Monitoring using Abstract Scene Graphs}
\label{sec:rmasg}
\begin{figure}[t]
	\centering
	\includegraphics[width=0.8\linewidth]{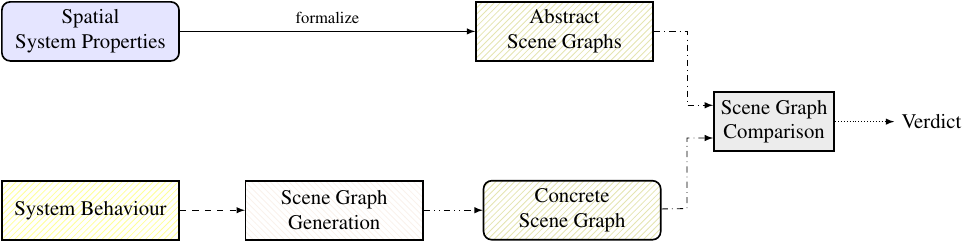}
	\caption{Runtime Monitoring using Abstract Scene Graphs}
    \label{fig:rm}
\end{figure}

\newcommand{\mon} { \mathrm{Mon} }
Runtime Monitoring~(RM) is a technique that is used to continuously check whether a system is satisfying its system properties.
It makes use of runtime monitors, which are software components that take system behaviour as input and provide an immediate verdict regarding satisfaction with the system's properties.
In this work, we focus on monitoring whether Automated Driving Functions~(ADFs) interacting in automotive traffic scenes are satisfying their spatial system properties during their operation.
As discussed before, such spatial properties need to be formalized, and we propose to use the Abstract Scene Graphs~(ASGs) formalism for this purpose.
\reffig{rm} shows an overview of our proposal for performing RM using ASGs.

We propose to model the required ADF system behaviour as a concrete traffic scene, since a concrete traffic scene contains information about all relevant traffic participants and traffic environment with which the ADF interacts.
A Scene Graph describing the concrete traffic scene is called a Concrete Scene Graph~(CSG). 
It is represented as $CSG: G_\text{C}$, where $G_\text{C}$ is a directed heterogeneous graph.
Semantically, a $G_\text{C}$ represents the traffic participant and traffic environment entities present in the traffic scene, their spatial relationships with each other and their corresponding attribute values at the current instance of time.  
In literature, a wide range of Scene Graph Generation~(SGG) techniques exist that help generate Scene Graphs based on continuously captured data from vehicle sensors, traffic infrastructure, simulation, etc.~\cite{Malawade2022, Yang2022, Woodlief2025b}.
During the SGG of a CSG, each identified object present in the traffic scene is assigned an object class~($\in\mathcal{C}$), as defined in the Object Model $OM$. 
Further, their attribute values, if available, are stored and spatial relations ($\in\mathcal{E}$) between pairs of identified objects are checked for existence. 

Finally, a mechanism is needed to check whether an ADF (whose behaviour is captured as a CSG) is satisfying the spatial properties as specified as ASGs.
In the following, we present one way in which a satisfaction relation between CSG and ASG can be defined.
Since both ASG $(G_{\text{A}}, D)$ and CSG consist of directed heterogeneous graphs, we propose to first find the sub-graph~($G'_\text{C}$) of $G_\text{C}$ that matches with $G_\text{A}$ in terms of object and relation types, i.e. $G'_\text{C}$ is isomorphic to $G_\text{A}$ or $G'_\text{C}\cong G_\text{A}$. 
We place an additional constraint here, that the node representing the ADF (usually referred to as ego) must be present in $G'_\text{C}$. 
Once a suitable subgraph ($G'_\text{C}$) is found, it is then checked whether the object attribute values stored in $G'_\text{C}$ satisfy the predicates present in $D$. 
Let $\varrho_\text{C}$ be the set of object attribute values stored in $G'_\text{C}$.
Based on this discussion, we define that a CSG ($G_\text{C}$) satisfies the ASG ($G_\text{A},D$) if
\begin{equation*}
	\text{CSG}\vDash \text{ASG}\Longleftrightarrow (\exists G'_\text{C}: G'_\text{C}\cong G_\text{A})\land (\varrho_\text{C}\vDash D)
\end{equation*}
We have implemented the above-mentioned process of checking the satisfaction relation between an ASG and CSG as a proof-of-concept algorithm as shown in~\refalg{sgc}.
The comparison algorithm expects a CSG $: G_\text{C}$ and an ASG $: (G_\text{A}, D)$ as inputs and provides a boolean verdict for whether the CSG satisfies the provided ASG. 
In the first step, the VF2 algorithm~\cite{Cordella2004} is used to identify the subgraph $G'_\text{C}$ of $G_\text{C}$ that is isomorphic to $G_\text{A}$. 
If $G'_\text{C}$ is found, a mapping between matched elements of $G_\text{C}$ and $G_\text{A}$ is also returned by the VF2 algorithm.
Given that a $G'_\text{C}$ exists, attribute values corresponding to those present in the Object Model are extracted from $G_\text{C}$.
Finally, it is checked whether the extracted attribute values satisfy the predicates present in $D$.
A $True$ verdict implies that the CSG satisfies the spatial properties specified as an ASG, while a $False$ verdict implies a violation of the properties. 
While implementing the algorithm, we assume that only a maximum of one subgraph of $G_\text{C}$ will be isomorphic to $G_\text{A}$. 
We realize that the use of subgraph isomorphism, which is a NP-complete problem, limits the applicability of the currently proposed Scene Graph comparison method for runtime applications and further work is required to find a suitable alternative. 

\begin{algorithm}
	\caption{Scene Graph Comparison}
	\small
\begin{algorithmic}
	\State \textbf{Input: }ASG $: (G_\text{A}, D)$ and CSG$: G_\text{C}$ \textbf{Output: }Boolean verdict 
\Procedure{sg_comparison}{$G_\text{A}, D$, $G_\text{C}$}
\State $(exists$, $map) \gets$ \Call{check_isomorphism_VF2}{$G_\text{A}$, $G_\text{C}$}
\If{$exists$ is $True$}
\State $attr\gets$\Call{get_mapped_attribute_values}{$map$, $G_\text{A}$, $G_\text{C}$}
\ForAll{$predicate\in D$}  
\State $pred\_verdict\gets$\Call{check_if_predicate_satisfied}{$attr$, $predicate$}
\If{$pred\_verdict$ is $False$} 
\State $\text{verdict}\gets False$ \Comment{If a predicate is not satisfied, CSG does not satisfy ASG}
\State \textbf{end procedure}
\EndIf
\EndFor
\State $\text{verdict}\gets True$\Comment{If all predicates are satisfied, CSG satisfies ASG}
\Else
\State $\text{verdict}\gets False$\Comment{No isomorphic subgraph exists}
\EndIf
\State \textbf{return} $\text{verdict}$
\EndProcedure
\end{algorithmic}
\label{alg:sgc}
\end{algorithm}

%% file: sections/conclusion.tex
\section{Conclusion and Future Work}
\label{sec:conclusion}
In this work, we presented a research preview on the Abstract Scene Graph~(ASG) formalism.
By using examples from the NHTSA pre-crash scenario catalogue, we showed how ASGs enable formalization of spatial properties of Automated Driving Functions~(ADFs) present in the automotive domain.
Finally, we also presented a framework for performing Runtime Monitoring~(RM) thereby checking the behaviour of ADFs against spatial properties formalized using ASGs.

As part of future work, we aim to extend the syntax of the ASG formalism by mapping its Object Model and allowed relationships between objects to a suitable ontology from the automotive domain. 
This would help improve the expressivity of ASGs and enable formalization of more varied and complex spatial properties. 
Further testing is also required regarding efficiency of the currently proposed framework for performing RM using ASGs, in particular, on automotive grade hardware. 
In addition, the Scene Graph comparison mechanism needs to be further developed and tested for more complex ASGs and Concrete Scene Graphs. 
Finally, applicability of the ASG formalism for specifying spatial properties in other domains such as indoor and maritime should also be checked.

%% file: sections/appendix.tex
\section*{Appendix}
\begin{figure}[h]
	\centering
	\includegraphics[scale=0.6]{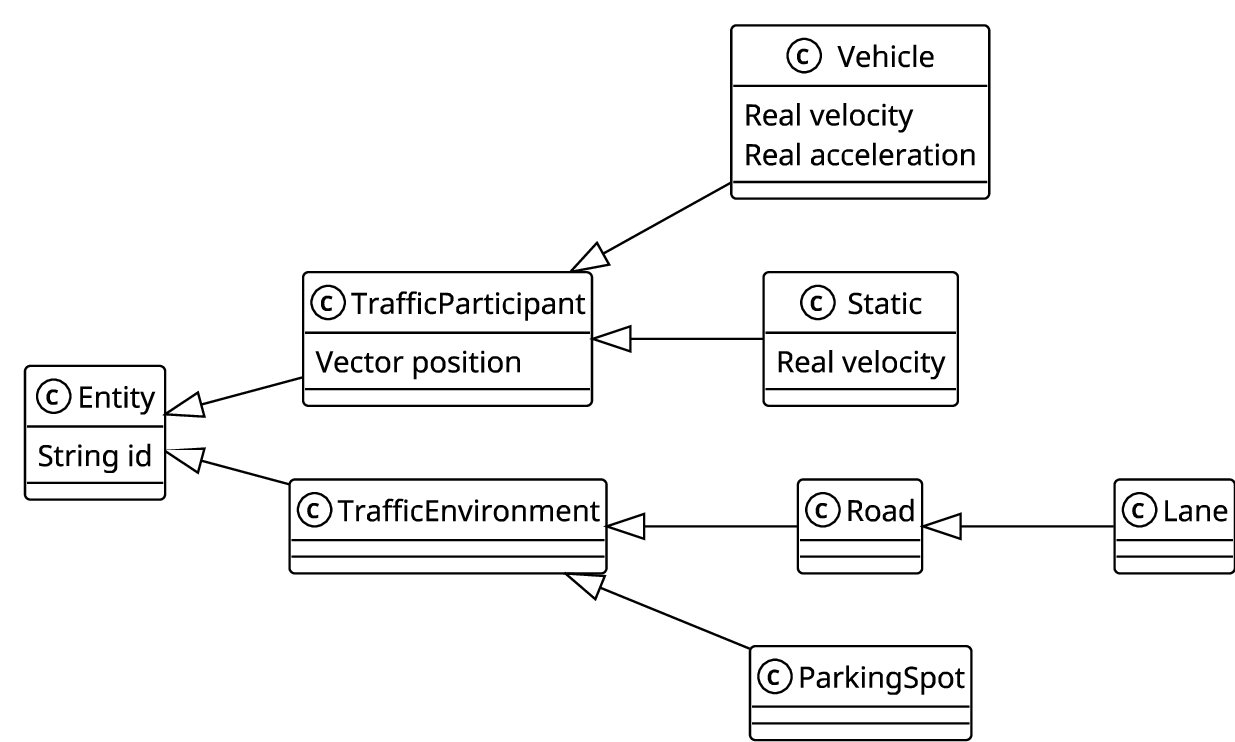}
	\caption{Object Model used for formalizing spatial properties as Abstract Scene Graphs}
	\label{fig:objectmodel}
\end{figure}

\begin{table}[h]
\caption{Allowed relationship types between various object types}
\label{tab:my-table}
\begin{tabular}{@{}c c c c l@{}}
\toprule
\textbf{Syntax} & \textbf{Relationship Type} & \textbf{Semantics} &  \\ \midrule
 isIn & Entity -> Lane & Entity is present in a lane &  \\ 
 isPartOf & Lane -> Road & Lane is part of a road &  \\ 
 inFrontOf & TrafficParticipant -> TrafficParticipant & Object (left) is present spatially in front of other Object &  \\ \bottomrule
\end{tabular}%
\end{table}
\begin{figure}[h]
\begin{subfigure}{.5\textwidth}
  \centering
  \includegraphics[width=\linewidth]{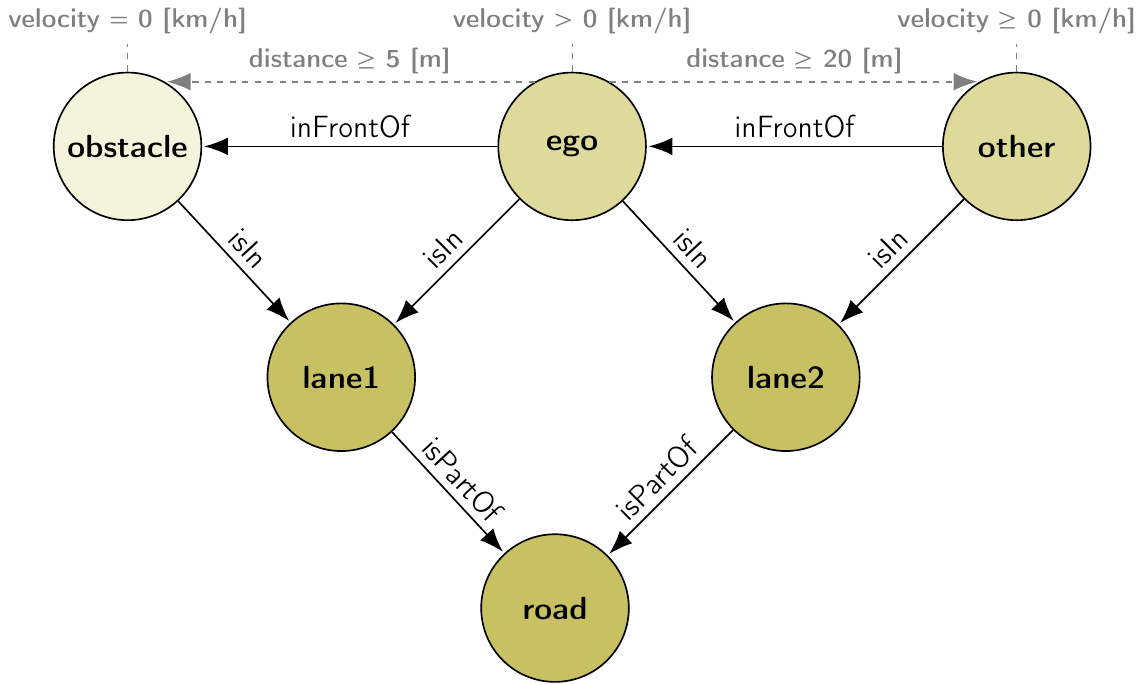}
  \caption{P2-4}
  \label{fig:sfig1}
\end{subfigure}%
\begin{subfigure}{.5\textwidth}
  \centering
  \includegraphics[width=.8\linewidth]{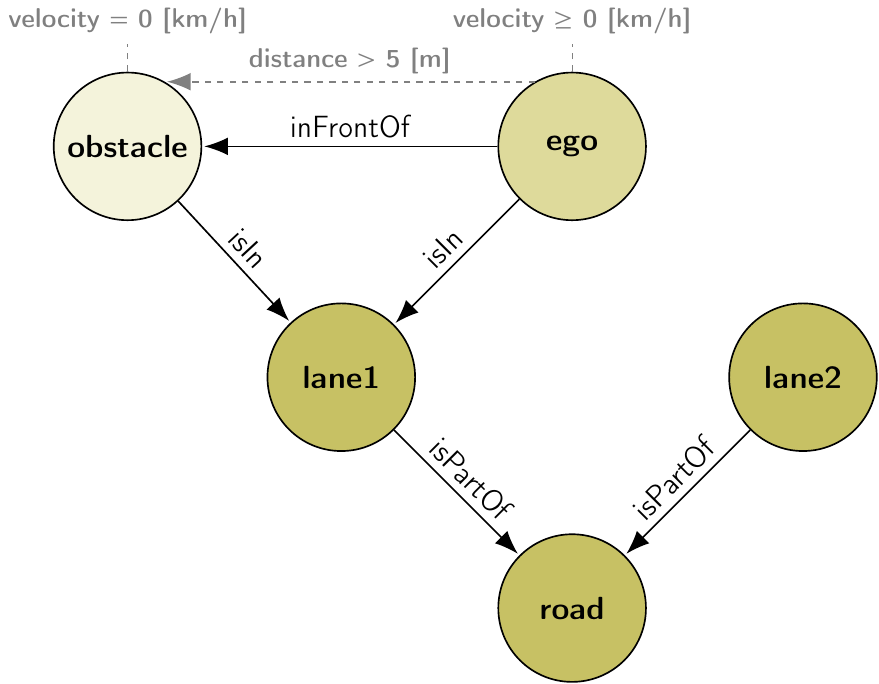}
  \caption{P2-5}
  \label{fig:sfig2}
\end{subfigure}
\caption{Remaining Abstract Scene Graphs for property \textbf{P2}}
\label{fig:fig}
\end{figure}

%% file: main.bbl
\begin{thebibliography}{10}
\providecommand{\bibitemdeclare}[2]{}
\providecommand{\surnamestart}{}
\providecommand{\surnameend}{}
\providecommand{\urlprefix}{Available at }
\providecommand{\url}[1]{\texttt{#1}}
\providecommand{\href}[2]{\texttt{#2}}
\providecommand{\urlalt}[2]{\href{#1}{#2}}
\providecommand{\doi}[1]{doi:\urlalt{https://doi.org/#1}{#1}}
\providecommand{\eprint}[1]{arXiv:\urlalt{https://arxiv.org/abs/#1}{#1}}
\providecommand{\bibinfo}[2]{#2}

\bibitemdeclare{article}{Cordella2004}
\bibitem{Cordella2004}
\bibinfo{author}{L.P. \surnamestart Cordella\surnameend},
  \bibinfo{author}{P.~\surnamestart Foggia\surnameend},
  \bibinfo{author}{C.~\surnamestart Sansone\surnameend} \&
  \bibinfo{author}{M.~\surnamestart Vento\surnameend} (\bibinfo{year}{2004}):
  \emph{\bibinfo{title}{A (sub)graph isomorphism algorithm for matching large
  graphs}}.
\newblock {\slshape \bibinfo{journal}{IEEE Transactions on Pattern Analysis and
  Machine Intelligence}} \bibinfo{volume}{26}(\bibinfo{number}{10}), pp.
  \bibinfo{pages}{1367--1372}, \doi{10.1109/TPAMI.2004.75}.

\bibitemdeclare{inproceedings}{Damm2018}
\bibitem{Damm2018}
\bibinfo{author}{Werner \surnamestart Damm\surnameend},
  \bibinfo{author}{Stefanie \surnamestart Kemper\surnameend},
  \bibinfo{author}{Eike \surnamestart Möhlmann\surnameend},
  \bibinfo{author}{Thomas \surnamestart Peikenkamp\surnameend} \&
  \bibinfo{author}{Astrid \surnamestart Rakow\surnameend}
  (\bibinfo{year}{2018}): \emph{\bibinfo{title}{Using {Traffic} {Sequence}
  {Charts} for the {Development} of {HAVs}}}.
\newblock In: {\slshape \bibinfo{booktitle}{{ERTS} 2018}}, \bibinfo{series}{9th
  {European} {Congress} on {Embedded} {Real} {Time} {Software} and {Systems}
  ({ERTS} 2018)}, \bibinfo{address}{Toulouse, France}.
\newblock \urlprefix\url{https://hal.science/hal-01714060}.

\bibitemdeclare{incollection}{Damm2018a}
\bibitem{Damm2018a}
\bibinfo{author}{Werner \surnamestart Damm\surnameend}, \bibinfo{author}{Eike
  \surnamestart Möhlmann\surnameend}, \bibinfo{author}{Thomas \surnamestart
  Peikenkamp\surnameend} \& \bibinfo{author}{Astrid \surnamestart
  Rakow\surnameend} (\bibinfo{year}{2018}): \emph{\bibinfo{title}{A {Formal}
  {Semantics} for {Traffic} {Sequence} {Charts}}}.
\newblock \bibinfo{series}{Lecture {Notes} in {Computer} {Science}},
  \bibinfo{publisher}{Springer International Publishing},
  \bibinfo{address}{Cham}, pp. \bibinfo{pages}{182--205},
  \doi{10.1007/978-3-319-95246-8_11}.

\bibitemdeclare{inproceedings}{Grundt2022}
\bibitem{Grundt2022}
\bibinfo{author}{Dominik \surnamestart Grundt\surnameend},
  \bibinfo{author}{Anna \surnamestart K\"ohne\surnameend},
  \bibinfo{author}{Ishan \surnamestart Saxena\surnameend},
  \bibinfo{author}{Ralf \surnamestart Stemmer\surnameend},
  \bibinfo{author}{Bernd \surnamestart Westphal\surnameend} \&
  \bibinfo{author}{Eike \surnamestart M\"ohlmann\surnameend}
  (\bibinfo{year}{2022}): \emph{\bibinfo{title}{Towards Runtime Monitoring of
  Complex System Requirements for Autonomous Driving Functions}}.
\newblock In: {\slshape \bibinfo{booktitle}{{\rm Proceedings Fourth
  International Workshop on} Formal Methods for Autonomous Systems (FMAS) {\rm
  and Fourth International Workshop on} Automated and verifiable Software
  sYstem DEvelopment (ASYDE), {\rm Berlin, Germany, 26th and 27th of September
  2022}}}, {\slshape \bibinfo{series}{Electronic Proceedings in Theoretical
  Computer Science}} \bibinfo{volume}{371}, \bibinfo{publisher}{Open Publishing
  Association}, pp. \bibinfo{pages}{53--61}, \doi{10.4204/EPTCS.371.4}.

\bibitemdeclare{inproceedings}{Hilscher2011}
\bibitem{Hilscher2011}
\bibinfo{author}{Martin \surnamestart Hilscher\surnameend},
  \bibinfo{author}{Sven \surnamestart Linker\surnameend},
  \bibinfo{author}{Ernst-Rüdiger \surnamestart Olderog\surnameend} \&
  \bibinfo{author}{Anders~P. \surnamestart Ravn\surnameend}
  (\bibinfo{year}{2011}): \emph{\bibinfo{title}{An {Abstract} {Model} for
  {Proving} {Safety} of {Multi}-lane {Traffic} {Manoeuvres}}}.
\newblock In \bibinfo{editor}{Shengchao \surnamestart Qin\surnameend} \&
  \bibinfo{editor}{Zongyan \surnamestart Qiu\surnameend}, editors: {\slshape
  \bibinfo{booktitle}{Formal {Methods} and {Software} {Engineering}}},
  \bibinfo{series}{Lecture {Notes} in {Computer} {Science}},
  \bibinfo{publisher}{Springer}, \bibinfo{address}{Berlin, Heidelberg}, pp.
  \bibinfo{pages}{404--419}, \doi{10.1007/978-3-642-24559-6_28}.

\bibitemdeclare{article}{Kutz2003}
\bibitem{Kutz2003}
\bibinfo{author}{Oliver \surnamestart Kutz\surnameend}, \bibinfo{author}{Frank
  \surnamestart Wolter\surnameend}, \bibinfo{author}{Holger \surnamestart
  Sturm\surnameend}, \bibinfo{author}{Nobu-Yuki \surnamestart
  Suzuki\surnameend} \& \bibinfo{author}{Michael \surnamestart
  Zakharyaschev\surnameend} (\bibinfo{year}{2003}):
  \emph{\bibinfo{title}{Logics of metric spaces}}.
\newblock {\slshape \bibinfo{journal}{ACM Transactions on Computational Logic}}
  \bibinfo{volume}{4}(\bibinfo{number}{2}), pp. \bibinfo{pages}{260--294},
  \doi{10.1145/635499.635504}.

\bibitemdeclare{article}{Li2021}
\bibitem{Li2021}
\bibinfo{author}{Jinghang \surnamestart Li\surnameend}, \bibinfo{author}{Chao
  \surnamestart Lu\surnameend}, \bibinfo{author}{Penghui \surnamestart
  Li\surnameend}, \bibinfo{author}{Zheyu \surnamestart Zhang\surnameend},
  \bibinfo{author}{Cheng \surnamestart Gong\surnameend} \&
  \bibinfo{author}{Jianwei \surnamestart Gong\surnameend}
  (\bibinfo{year}{2021}): \emph{\bibinfo{title}{Driver-Specific Risk
  Recognition in Interactive Driving Scenarios using Graph Representation}}.
\newblock \doi{10.48550/ARXIV.2111.06342}.

\bibitemdeclare{inproceedings}{Li2020}
\bibitem{Li2020}
\bibinfo{author}{Tengfei \surnamestart Li\surnameend}, \bibinfo{author}{Jing
  \surnamestart Liu\surnameend}, \bibinfo{author}{JieXiang \surnamestart
  Kang\surnameend}, \bibinfo{author}{Haiying \surnamestart Sun\surnameend},
  \bibinfo{author}{Wei \surnamestart Yin\surnameend}, \bibinfo{author}{Xiaohong
  \surnamestart Chen\surnameend} \& \bibinfo{author}{Hui \surnamestart
  Wang\surnameend} (\bibinfo{year}{2020}): \emph{\bibinfo{title}{{STSL}: {A}
  {Novel} {Spatio}-{Temporal} {Specification} {Language} for {Cyber}-{Physical}
  {Systems}}}.
\newblock In: {\slshape \bibinfo{booktitle}{2020 {IEEE} 20th {International}
  {Conference} on {Software} {Quality}, {Reliability} and {Security} ({QRS})}},
  pp. \bibinfo{pages}{309--319}, \doi{10.1109/QRS51102.2020.00048}.

\bibitemdeclare{article}{Li2021a}
\bibitem{Li2021a}
\bibinfo{author}{Tengfei \surnamestart Li\surnameend}, \bibinfo{author}{Jing
  \surnamestart Liu\surnameend}, \bibinfo{author}{Haiying \surnamestart
  Sun\surnameend}, \bibinfo{author}{Xiaohong \surnamestart Chen\surnameend},
  \bibinfo{author}{Ling \surnamestart Yin\surnameend}, \bibinfo{author}{Xia
  \surnamestart Mao\surnameend} \& \bibinfo{author}{Junfeng \surnamestart
  Sun\surnameend} (\bibinfo{year}{2021}): \emph{\bibinfo{title}{Runtime
  {Verification} of {Spatio}-{Temporal} {Specification} {Language}}}.
\newblock {\slshape \bibinfo{journal}{Mobile Networks and Applications}}
  \bibinfo{volume}{26}(\bibinfo{number}{6}), pp. \bibinfo{pages}{2392--2406},
  \doi{10.1007/s11036-021-01779-5}.

\bibitemdeclare{article}{Li2023}
\bibitem{Li2023}
\bibinfo{author}{Tianyu \surnamestart Li\surnameend},
  \bibinfo{author}{Li~\surnamestart Chen\surnameend}, \bibinfo{author}{Huijie
  \surnamestart Wang\surnameend}, \bibinfo{author}{Yang \surnamestart
  Li\surnameend}, \bibinfo{author}{Jiazhi \surnamestart Yang\surnameend},
  \bibinfo{author}{Xiangwei \surnamestart Geng\surnameend},
  \bibinfo{author}{Shengyin \surnamestart Jiang\surnameend},
  \bibinfo{author}{Yuting \surnamestart Wang\surnameend}, \bibinfo{author}{Hang
  \surnamestart Xu\surnameend}, \bibinfo{author}{Chunjing \surnamestart
  Xu\surnameend}, \bibinfo{author}{Junchi \surnamestart Yan\surnameend},
  \bibinfo{author}{Ping \surnamestart Luo\surnameend} \&
  \bibinfo{author}{Hongyang \surnamestart Li\surnameend}
  (\bibinfo{year}{2023}): \emph{\bibinfo{title}{Graph-based Topology Reasoning
  for Driving Scenes}}.
\newblock \doi{10.48550/ARXIV.2304.05277}.

\bibitemdeclare{article}{Malawade2022}
\bibitem{Malawade2022}
\bibinfo{author}{Arnav~Vaibhav \surnamestart Malawade\surnameend},
  \bibinfo{author}{Shih-Yuan \surnamestart Yu\surnameend},
  \bibinfo{author}{Brandon \surnamestart Hsu\surnameend},
  \bibinfo{author}{Harsimrat \surnamestart Kaeley\surnameend},
  \bibinfo{author}{Anurag \surnamestart Karra\surnameend} \&
  \bibinfo{author}{Mohammad~Abdullah \surnamestart {Al Faruque}\surnameend}
  (\bibinfo{year}{2022}): \emph{\bibinfo{title}{roadscene2vec: A tool for
  extracting and embedding road scene-graphs}}.
\newblock {\slshape \bibinfo{journal}{Knowledge-Based Systems}}
  \bibinfo{volume}{242}, p. \bibinfo{pages}{108245},
  \doi{10.1016/j.knosys.2022.108245}.

\bibitemdeclare{article}{Pedro2024}
\bibitem{Pedro2024}
\bibinfo{author}{André \surnamestart Matos Pedro\surnameend},
  \bibinfo{author}{Tomás \surnamestart Silva\surnameend},
  \bibinfo{author}{Tiago \surnamestart Sequeira\surnameend},
  \bibinfo{author}{João \surnamestart Lourenço\surnameend},
  \bibinfo{author}{João~Costa \surnamestart Seco\surnameend} \&
  \bibinfo{author}{Carla \surnamestart Ferreira\surnameend}
  (\bibinfo{year}{2024}): \emph{\bibinfo{title}{Monitoring of spatio-temporal
  properties with nonlinear {SAT} solvers}}.
\newblock {\slshape \bibinfo{journal}{International Journal on Software Tools
  for Technology Transfer}}, \doi{10.1007/s10009-024-00740-7}.

\bibitemdeclare{article}{Menzel2018}
\bibitem{Menzel2018}
\bibinfo{author}{Till \surnamestart Menzel\surnameend}, \bibinfo{author}{Gerrit
  \surnamestart Bagschik\surnameend} \& \bibinfo{author}{Markus \surnamestart
  Maurer\surnameend} (\bibinfo{year}{2018}): \emph{\bibinfo{title}{Scenarios
  for Development, Test and Validation of Automated Vehicles}}.
\newblock \doi{10.48550/ARXIV.1801.08598}.

\bibitemdeclare{article}{Monninger2023}
\bibitem{Monninger2023}
\bibinfo{author}{Thomas \surnamestart Monninger\surnameend},
  \bibinfo{author}{Julian \surnamestart Schmidt\surnameend},
  \bibinfo{author}{Jan \surnamestart Rupprecht\surnameend},
  \bibinfo{author}{David \surnamestart Raba\surnameend},
  \bibinfo{author}{Julian \surnamestart Jordan\surnameend},
  \bibinfo{author}{Daniel \surnamestart Frank\surnameend},
  \bibinfo{author}{Steffen \surnamestart Staab\surnameend} \&
  \bibinfo{author}{Klaus \surnamestart Dietmayer\surnameend}
  (\bibinfo{year}{2023}): \emph{\bibinfo{title}{SCENE: Reasoning about Traffic
  Scenes using Heterogeneous Graph Neural Networks}}.
\newblock {\slshape \bibinfo{journal}{IEEE Robotics and Automation Letters
  (RA-L), 2023}} \bibinfo{volume}{8}(\bibinfo{number}{3}), pp.
  \bibinfo{pages}{1531--1538}, \doi{10.1109/lra.2023.3234771}.

\bibitemdeclare{techreport}{Najm2007}
\bibitem{Najm2007}
\bibinfo{author}{Wassim~G \surnamestart Najm\surnameend},
  \bibinfo{author}{John~D. \surnamestart Smith\surnameend} \&
  \bibinfo{author}{Mikio \surnamestart Yanagisawa\surnameend}
  (\bibinfo{year}{2007}): \emph{\bibinfo{title}{Pre-crash scenario typology for
  crash avoidance research}}.
\newblock \bibinfo{type}{Technical Report}
  \bibinfo{number}{DOT-VNTSC-NHTSA-06-02}.
\newblock \urlprefix\url{https://rosap.ntl.bts.gov/view/dot/6281}.

\bibitemdeclare{inproceedings}{Neurohr2020}
\bibitem{Neurohr2020}
\bibinfo{author}{Christian \surnamestart Neurohr\surnameend},
  \bibinfo{author}{Lukas \surnamestart Westhofen\surnameend},
  \bibinfo{author}{Tabea \surnamestart Henning\surnameend},
  \bibinfo{author}{Thies \surnamestart de~Graaff\surnameend},
  \bibinfo{author}{Eike \surnamestart Möhlmann\surnameend} \&
  \bibinfo{author}{Eckard \surnamestart Böde\surnameend}
  (\bibinfo{year}{2020}): \emph{\bibinfo{title}{Fundamental {Considerations}
  around {Scenario}-{Based} {Testing} for {Automated} {Driving}}}.
\newblock In: {\slshape \bibinfo{booktitle}{2020 {IEEE} {Intelligent}
  {Vehicles} {Symposium} ({IV})}}, pp. \bibinfo{pages}{121--127},
  \doi{10.1109/IV47402.2020.9304823}.

\bibitemdeclare{misc}{Neurohr2025}
\bibitem{Neurohr2025}
\bibinfo{author}{Christian \surnamestart Neurohr\surnameend},
  \bibinfo{author}{Lukas \surnamestart Westhofen\surnameend},
  \bibinfo{author}{Tjark \surnamestart Koopmann\surnameend},
  \bibinfo{author}{Eike \surnamestart Möhlmann\surnameend},
  \bibinfo{author}{Eckard \surnamestart Böde\surnameend} \&
  \bibinfo{author}{Axel \surnamestart Hahn\surnameend} (\bibinfo{year}{2025}):
  \emph{\bibinfo{title}{On Scenario Formalisms for Automated Driving}},
  \doi{10.48550/arXiv.2504.04868}.

\bibitemdeclare{inproceedings}{Ody2017}
\bibitem{Ody2017}
\bibinfo{author}{Heinrich \surnamestart Ody\surnameend} (\bibinfo{year}{2017}):
  \emph{\bibinfo{title}{Monitoring of Traffic Manoeuvres with Imprecise
  Information}}.
\newblock In \bibinfo{editor}{Lukas \surnamestart Bulwahn\surnameend},
  \bibinfo{editor}{Maryam \surnamestart Kamali\surnameend} \&
  \bibinfo{editor}{Sven \surnamestart Linker\surnameend}, editors: {\slshape
  \bibinfo{booktitle}{{Proceedings First Workshop on} Formal Verification of
  Autonomous Vehicles, {Turin, Italy, 19th September 2017}}}, {\slshape
  \bibinfo{series}{Electronic Proceedings in Theoretical Computer Science}}
  \bibinfo{volume}{257}, \bibinfo{publisher}{Open Publishing Association}, pp.
  \bibinfo{pages}{43--58}, \doi{10.4204/EPTCS.257.6}.

\bibitemdeclare{inproceedings}{Saxena2025}
\bibitem{Saxena2025}
\bibinfo{author}{Ishan \surnamestart Saxena\surnameend}, \bibinfo{author}{Bernd
  \surnamestart Westphal\surnameend} \& \bibinfo{author}{Martin \surnamestart
  Fränzle\surnameend} (\bibinfo{year}{2025}): \emph{\bibinfo{title}{Towards
  Runtime Monitoring of spatial properties of Automated Driving Functions using
  Abstract Scene Graphs}}.
\newblock In: {\slshape \bibinfo{booktitle}{Young Researchers Seminar 2025 -
  Proceedings of the 12th Young Researchers Seminar, hosted by BASt, 3-5 June,
  2025 (YRS 2025)}}, \doi{10.60850/bericht-a53}.

\bibitemdeclare{inproceedings}{Schwammberger2021}
\bibitem{Schwammberger2021}
\bibinfo{author}{Maike \surnamestart Schwammberger\surnameend} \&
  \bibinfo{author}{Gleifer \surnamestart Vaz~Alves\surnameend}
  (\bibinfo{year}{2021}): \emph{\bibinfo{title}{Extending {Urban}
  {Multi}-{Lane} {Spatial} {Logic} to {Formalise} {Road} {Junction} {Rules}}}.
\newblock In: {\slshape \bibinfo{booktitle}{Proceedings {Third} {Workshop} on
  {Formal} {Methods} for {Autonomous} {Systems}, {FMAS} 2021, {Virtual},
  21st-22nd of {October} 2021, {Ed}.: {M}. {Farrell}}}, p.~\bibinfo{pages}{1},
  \doi{10.4204/EPTCS.348.1}.

\bibitemdeclare{article}{Stemmer2025}
\bibitem{Stemmer2025}
\bibinfo{author}{Ralf \surnamestart Stemmer\surnameend}, \bibinfo{author}{Ishan
  \surnamestart Saxena\surnameend}, \bibinfo{author}{Lukas \surnamestart
  Panneke\surnameend}, \bibinfo{author}{Dominik \surnamestart
  Grundt\surnameend}, \bibinfo{author}{Anna \surnamestart Austel\surnameend},
  \bibinfo{author}{Eike \surnamestart Möhlmann\surnameend} \&
  \bibinfo{author}{Bernd \surnamestart Westphal\surnameend}
  (\bibinfo{year}{2025}): \emph{\bibinfo{title}{Runtime monitoring of complex
  scenario-based requirements for autonomous driving functions}}.
\newblock {\slshape \bibinfo{journal}{Science of Computer Programming}}
  \bibinfo{volume}{244}, p. \bibinfo{pages}{103301},
  \doi{10.1016/j.scico.2025.103301}.

\bibitemdeclare{inproceedings}{Toledo2024}
\bibitem{Toledo2024}
\bibinfo{author}{Felipe \surnamestart Toledo\surnameend}, \bibinfo{author}{Trey
  \surnamestart Woodlief\surnameend}, \bibinfo{author}{Sebastian \surnamestart
  Elbaum\surnameend} \& \bibinfo{author}{Matthew~B. \surnamestart
  Dwyer\surnameend} (\bibinfo{year}{2024}): \emph{\bibinfo{title}{Specifying
  and Monitoring Safe Driving Properties with Scene Graphs}}.
\newblock In: {\slshape \bibinfo{booktitle}{2024 {IEEE} {International}
  {Conference} on {Robotics} and {Automation} ({ICRA})}}, pp.
  \bibinfo{pages}{15577--15584}, \doi{10.1109/ICRA57147.2024.10610973}.

\bibitemdeclare{inproceedings}{Ulbrich2015}
\bibitem{Ulbrich2015}
\bibinfo{author}{Simon \surnamestart Ulbrich\surnameend}, \bibinfo{author}{Till
  \surnamestart Menzel\surnameend}, \bibinfo{author}{Andreas \surnamestart
  Reschka\surnameend}, \bibinfo{author}{Fabian \surnamestart
  Schuldt\surnameend} \& \bibinfo{author}{Markus \surnamestart
  Maurer\surnameend} (\bibinfo{year}{2015}): \emph{\bibinfo{title}{Defining and
  Substantiating the Terms Scene, Situation, and Scenario for Automated
  Driving}}.
\newblock In: {\slshape \bibinfo{booktitle}{2015 IEEE 18th International
  Conference on Intelligent Transportation Systems}}, pp.
  \bibinfo{pages}{982--988}, \doi{10.1109/ITSC.2015.164}.

\bibitemdeclare{article}{Wang2023}
\bibitem{Wang2023}
\bibinfo{author}{Junyao \surnamestart Wang\surnameend},
  \bibinfo{author}{Arnav~Vaibhav \surnamestart Malawade\surnameend},
  \bibinfo{author}{Junhong \surnamestart Zhou\surnameend},
  \bibinfo{author}{Shih-Yuan \surnamestart Yu\surnameend} \&
  \bibinfo{author}{Mohammad Abdullah~Al \surnamestart Faruque\surnameend}
  (\bibinfo{year}{2023}): \emph{\bibinfo{title}{RS2G: Data-Driven Scene-Graph
  Extraction and Embedding for Robust Autonomous Perception and Scenario
  Understanding}}.
\newblock \doi{10.48550/ARXIV.2304.08600}.

\bibitemdeclare{article}{Woodlief2025a}
\bibitem{Woodlief2025a}
\bibinfo{author}{Trey \surnamestart Woodlief\surnameend},
  \bibinfo{author}{Felipe \surnamestart Toledo\surnameend},
  \bibinfo{author}{Matthew \surnamestart Dwyer\surnameend} \&
  \bibinfo{author}{Sebastian \surnamestart Elbaum\surnameend}
  (\bibinfo{year}{2025}): \emph{\bibinfo{title}{Scene {Flow} {Specifications}:
  {Encoding} and {Monitoring} {Rich} {Temporal} {Safety} {Properties} of
  {Autonomous} {Systems}}}.
\newblock {\slshape \bibinfo{journal}{Proc. ACM Softw. Eng.}}
  \bibinfo{volume}{2}(\bibinfo{number}{FSE}), pp.
  \bibinfo{pages}{FSE112:2524--FSE112:2547}, \doi{10.1145/3729382}.

\bibitemdeclare{inproceedings}{Woodlief2024}
\bibitem{Woodlief2024}
\bibinfo{author}{Trey \surnamestart Woodlief\surnameend},
  \bibinfo{author}{Felipe \surnamestart Toledo\surnameend},
  \bibinfo{author}{Sebastian \surnamestart Elbaum\surnameend} \&
  \bibinfo{author}{Matthew~B \surnamestart Dwyer\surnameend}
  (\bibinfo{year}{2024}): \emph{\bibinfo{title}{{S3C}: {Spatial} {Semantic}
  {Scene} {Coverage} for {Autonomous} {Vehicles}}}.
\newblock In: {\slshape \bibinfo{booktitle}{Proceedings of the {IEEE}/{ACM}
  46th {International} {Conference} on {Software} {Engineering}}},
  \bibinfo{series}{{ICSE} '24}, \bibinfo{publisher}{Association for Computing
  Machinery}, \bibinfo{address}{New York, NY, USA}, pp. \bibinfo{pages}{1--13},
  \doi{10.1145/3597503.3639178}.

\bibitemdeclare{inproceedings}{Woodlief2025b}
\bibitem{Woodlief2025b}
\bibinfo{author}{Trey \surnamestart Woodlief\surnameend},
  \bibinfo{author}{Felipe \surnamestart Toledo\surnameend},
  \bibinfo{author}{Sebastian \surnamestart Elbaum\surnameend} \&
  \bibinfo{author}{Matthew~B. \surnamestart Dwyer\surnameend}
  (\bibinfo{year}{2025}): \emph{\bibinfo{title}{Closing the Gap Between Sensor
  Inputs and Driving Properties: A Scene Graph Generator for CARLA}}.
\newblock In: {\slshape \bibinfo{booktitle}{2025 IEEE/ACM 47th International
  Conference on Software Engineering: Companion Proceedings (ICSE-Companion)}},
  pp. \bibinfo{pages}{29--32}, \doi{10.1109/ICSE-Companion66252.2025.00017}.

\bibitemdeclare{inproceedings}{Yang2022}
\bibitem{Yang2022}
\bibinfo{author}{Jingkang \surnamestart Yang\surnameend},
  \bibinfo{author}{Yi~Zhe \surnamestart Ang\surnameend}, \bibinfo{author}{Zujin
  \surnamestart Guo\surnameend}, \bibinfo{author}{Kaiyang \surnamestart
  Zhou\surnameend}, \bibinfo{author}{Wayne \surnamestart Zhang\surnameend} \&
  \bibinfo{author}{Ziwei \surnamestart Liu\surnameend} (\bibinfo{year}{2022}):
  \emph{\bibinfo{title}{Panoptic Scene Graph Generation}}.
\newblock In \bibinfo{editor}{Shai \surnamestart Avidan\surnameend},
  \bibinfo{editor}{Gabriel \surnamestart Brostow\surnameend},
  \bibinfo{editor}{Moustapha \surnamestart Ciss{\'e}\surnameend},
  \bibinfo{editor}{Giovanni~Maria \surnamestart Farinella\surnameend} \&
  \bibinfo{editor}{Tal \surnamestart Hassner\surnameend}, editors: {\slshape
  \bibinfo{booktitle}{Computer Vision -- ECCV 2022}},
  \bibinfo{publisher}{Springer Nature Switzerland}, \bibinfo{address}{Cham},
  pp. \bibinfo{pages}{178--196}, \doi{10.1007/978-3-031-19812-0_11}.

\bibitemdeclare{article}{Yu2020}
\bibitem{Yu2020}
\bibinfo{author}{Shih-Yuan \surnamestart Yu\surnameend},
  \bibinfo{author}{Arnav~Vaibhav \surnamestart Malawade\surnameend},
  \bibinfo{author}{Deepan \surnamestart Muthirayan\surnameend},
  \bibinfo{author}{Pramod~P. \surnamestart Khargonekar\surnameend} \&
  \bibinfo{author}{Mohammad Abdullah~Al \surnamestart Faruque\surnameend}
  (\bibinfo{year}{2022}): \emph{\bibinfo{title}{Scene-Graph Augmented
  Data-Driven Risk Assessment of Autonomous Vehicle Decisions}}.
\newblock {\slshape \bibinfo{journal}{IEEE Transactions on Intelligent
  Transportation Systems}} \bibinfo{volume}{23}(\bibinfo{number}{7}), pp.
  \bibinfo{pages}{7941--7951}, \doi{10.1109/TITS.2021.3074854}.

\bibitemdeclare{article}{Zipfl2022}
\bibitem{Zipfl2022}
\bibinfo{author}{Maximilian \surnamestart Zipfl\surnameend},
  \bibinfo{author}{Felix \surnamestart Hertlein\surnameend},
  \bibinfo{author}{Achim \surnamestart Rettinger\surnameend},
  \bibinfo{author}{Steffen \surnamestart Thoma\surnameend},
  \bibinfo{author}{Lavdim \surnamestart Halilaj\surnameend},
  \bibinfo{author}{Juergen \surnamestart Luettin\surnameend},
  \bibinfo{author}{Stefan \surnamestart Schmid\surnameend} \&
  \bibinfo{author}{Cory \surnamestart Henson\surnameend}
  (\bibinfo{year}{2022}): \emph{\bibinfo{title}{Relation-based Motion
  Prediction using Traffic Scene Graphs}}.
\newblock \doi{10.1109/itsc55140.2022.9922155}.

\bibitemdeclare{misc}{Zipfl2023a}
\bibitem{Zipfl2023a}
\bibinfo{author}{Maximilian \surnamestart Zipfl\surnameend},
  \bibinfo{author}{Moritz \surnamestart Jarosch\surnameend} \&
  \bibinfo{author}{J.~Marius \surnamestart Zöllner\surnameend}
  (\bibinfo{year}{2023}): \emph{\bibinfo{title}{Traffic Scene Similarity: a
  Graph-based Contrastive Learning Approach}}, \doi{10.48550/ARXIV.2309.09720}.

\bibitemdeclare{article}{Zipfl2023}
\bibitem{Zipfl2023}
\bibinfo{author}{Maximilian \surnamestart Zipfl\surnameend},
  \bibinfo{author}{Sven \surnamestart Spickermann\surnameend} \&
  \bibinfo{author}{J.~Marius \surnamestart Zöllner\surnameend}
  (\bibinfo{year}{2023}): \emph{\bibinfo{title}{Utilizing Hybrid Trajectory
  Prediction Models to Recognize Highly Interactive Traffic Scenarios}}.
\newblock \doi{10.48550/ARXIV.2309.06887}.

\end{thebibliography}
